\def\rg{$r_{\rm g}$}
\def\phpspsqcm{ph\thinspace s$^{-1}$\thinspace cm$^{-2}$}
\def\cm{{\rm\thinspace cm}}

\def\erg{{\rm\thinspace erg}}
\def\eV{{\rm\thinspace eV}}

\def\Msun{\hbox{$\rm\thinspace M_{\odot}$}}
\def\Zsun{\hbox{$\rm\thinspace Z_{\odot}$}}

\def\s{{\rm\thinspace s}}

\def\ergpcmsqps{\hbox{$\erg\cm^{-2}\s^{-1}\,$}}

\def\psqcm{\hbox{$\cm^{-2}\,$}}

\def\spose#1{\hbox to 0pt{#1\hss}}
\def\approxlt{\mathrel{\spose{\lower 3pt\hbox{$\sim$}}
        \raise 2.0pt\hbox{$<$}}}
\def\approxgt{\mathrel{\spose{\lower 3pt\hbox{$\sim$}}
        \raise 2.0pt\hbox{$>$}}}
\def\rg{r_{\rm g}}
\def\rms{r_{\rm ms}}
\def\rin{r_{\rm in}}
\def\rout{r_{\rm out}}

\documentclass{mn2e}
\usepackage{times}

\input{psfig.sty}

\newif\ifAMStwofonts

\title[The continuum variability of MCG--6-30-15]
        {The continuum variability of MCG--6-30-15: A detailed analysis of the long 1999 \textit{ASCA} observation}
\author[D.~C.~Shih et al.]
        {D.~C.~Shih,$^{1,2}$ K.~Iwasawa$^2$ and A.~C.~Fabian$^2$
\\$1$ Department of Physics, Princeton University, Princeton, NJ 08544, USA.
\\$2$ Institute of Astronomy, Madingley Road, Cambridge, CB3 0HA.
}

\date{}

\begin{document}

\maketitle

\label{firstpage}

\begin{abstract}
We report on an analysis in the 3--10~keV X-ray band of the long 1999 \textit{ASCA} observation 
of MCG--6-30-15. The time-averaged broad iron K line is well-described by disk emission near a 
Schwarzschild black hole, confirming the results of earlier analyses on the \textit{ASCA} 1994 
and 1997 data. The time-resolved iron-line profile is remarkably stable over a factor of three 
change in source flux, and the line and continuum fluxes are uncorrelated. Detailed fits to
the variable iron-line profile suggest that the active region (parametrized by the best-fit inner
and outer radii of the accretion disk) responsible for iron line emission actually narrows with 
increasing flux to a region around 4--5 $\rg$. 
In contrast to the iron line, the power-law continuum exhibits significant 
variability during the 1999 observation. Time-resolved spectral 
analysis reveals a new feature in the well-known photon index ($\Gamma$) vs.\ flux correlation: 
$\Gamma$ appears to approach a limiting value of $\Gamma \sim 2.1$ at high flux. Two models are 
proposed to explain both the new feature in the $\Gamma$ vs.\ flux correlation and the 
uncorrelated iron-line flux: a phenomenological two power-law model, and the recently
proposed ``thundercloud'' model of Merloni \& Fabian (2001). Both models are capable of 
reproducing the data well, but because they are poorly constrained by the observed 
$\Gamma$ vs.\ flux relation, they cannot at present be tested meaningfully by the data. 
The various implications and the physical interpretation of these models are discussed.
\end{abstract}

\begin{keywords}
galaxies: individual: MCG--6-30-15 -- galaxies: Seyfert -- X-rays: galaxies.
\end{keywords}

\section[Introduction]
{Introduction}

In recent years, a self-consistent and robust picture 
has gradually emerged in order to explain the 
most important features of AGN X-ray spectra.
In the standard model, a central 
supermassive black hole accretes matter in the form of a cold, 
optically-thick disk. A hot, optically-thin corona above and below
the accretion disk inverse Compton scatters soft optical/UV 
photons from the disk, producing a power-law continuum in the 
X-ray band \cite{Zdziarski94}. At the same time, the corona illuminates 
the disk, giving rise to a reflection hump at higher 
energies and a prominent iron K line at around 6.4~keV 
\cite{Guilbert88,Lightman88}.

Extensive observations of MCG--6-30-15, a bright, 
nearby ($z=0.0078$) Seyfert 1, have been especially 
useful in the study of AGN X-ray spectra. In fact, MCG--6-30-15
is perhaps the most well-studied AGN in the X-ray band, owing to its
brightness, strong iron-line, and extreme variability. Because both
time-averaged and time-resolved spectral analysis are feasible for this 
source, its physical properties can be probed in great detail.
The many interesting features of the time-averaged spectrum,
such as the broad iron line, the reflection continuum, and the warm absorber,
have proven to be extremely useful 
in constraining the geometry and physics of AGN.

The study of spectral variability can provide another window into the
interesting properties of AGN, although it is generally much more difficult
and ambiguous than the simpler task of describing the time-averaged spectrum. 
Correlations (and lack thereof) have been observed between many 
different time-resolved spectral parameters,
and there is much debate over the theoretical interpretation. To mention two 
relevant examples, analyses by Lee et al.\ (2000) and 
Vaughan \& Edelson (2001) of a long 1997 \textit{RXTE} dataset
detected a steepening of the continuum spectrum with increasing 
flux in MCG--6-30-15. Meanwhile the iron line flux, though variable, 
did not appear to be correlated with the continuum flux. 

In this paper, we present an analysis of the \textit{ASCA} 1999 long 
observation of MCG--6-30-15 in the 3--10~keV X-ray band, 
investigating both the time-averaged properties of the spectrum and 
the time-resolved spectral variability. Section \ref{sec_obsanddatared} 
briefly describes the details of the 1999 observation and the status 
of the detector at the time. In Section \ref{sec_continuum_and_fka} the 
analysis of the time-averaged spectrum is outlined. The time-averaged 
spectrum shows the presence of a strong, relativistically-broadened 
iron line, confirming the results of earlier analyses on the 
1994 and 1997 \textit{ASCA} data. Section \ref{sec_tresspec} contains 
the results of the variability studies. The $\Gamma$ vs.\ flux 
relation is investigated, and a break in the usual correlation is observed.
Iron-line variability with respect to flux is also studied, and in particular,
the iron-line and continuum fluxes are found to be uncorrelated.
Possible interpretations of the flattening of $\Gamma$ are 
considered in Section \ref{sec_interpretation}, including an application
of the recently proposed ``thundercloud model'' of Merloni \& Fabian (2001).
Finally, Section \ref{sec_summary} summarizes the results of the analysis.

\section[Observations and Data Reduction]
{Observations and Data Reduction}
\label{sec_obsanddatared}

MCG--6-30-15 was observed with ASCA from 1999 July 19 to 1999 July 29.
The total integration time was $\sim 910$ ks. The data were
filtered and spectral files were prepared following the method of 
Iwasawa et al.\ (1999). 
The Solid-State Imaging 
Spectrometer (SIS0 and SIS1) was operated in Faint mode for the 
duration of the observation, and the Gas Imaging Spectrometer 
(GIS2 and GIS3) was operated in PH mode. Good exposure time 
amounted to approximately 440 ks for the SIS and 360 ks for the GIS. 
The FWHM energy resolutions 
at 6.4~keV for the SIS and GIS were $\sim$~350~eV and $\sim$~490~eV, 
respectively. This represents a degradation of a factor of 3--4 in SIS energy
resolution from launch, resulting from radiation damage to the CCD chips. 
The GIS detectors, on the other hand, being gas-based, are relatively
unaffected by the effects of high energy radiation, and so their
spectral resolution has remained very stable over time.

\section[Time-Averaged Spectral Features]
{Time-Averaged Spectral Features}
\label{sec_continuum_and_fka}

The total integrated count rates between 3--10~keV for the current 
observation were 0.540 and 0.518 cts/s for the SIS and GIS, respectively. 
The corresponding time-averaged flux in the 3--10~keV band was 
$3.20\times 10^{-11}\ergpcmsqps$. This represents a $20\%$ increase 
over the 3--10~keV flux in the previous 1997 observation, indicating
a significant brightening of the source between the two observations.

The 3--10~keV spectrum of MCG--6-30-15 was described using two {\sc xspec} model
components: {\tt pexrav} for the power-law continuum and cold reflection 
from the accretion disk \cite{Magdziarz95}; and {\tt diskline}
for the relativistically-broadened iron K$\alpha$ line at a rest-frame energy of 6.4~keV from
an accretion disk around a Schwarzschild black hole \cite{Fabian89}. 
In addition, {\tt pexrav} was relativistically blurred using
{\tt rdblur} \cite{Fabian89}, and the entire model was 
modified by cold Galactic absorption of $N_{\rm H} = 4.06\times 10^{20}$\psqcm \cite{Elvis89}.

The free parameters consisted of the inner and outer disk radius, 
disk inclination angle ({\tt diskline} and {\tt rdblur}); the total line flux ({\tt diskline});
and the power-law slope and normalization (defined to be the photon flux at 1~keV of the primary power law)
({\tt pexrav}).
The inclination angle relative to the observer of the reflecting material in {\tt pexrav} was
tied to the corresponding {\tt diskline} parameter. Since the other {\tt pexrav} parameters affect 
primarily the high energy spectrum, which is inaccessible
to \textit{ASCA}, they could not be constrained in this analysis, and so they were kept
fixed at values consistent with recent measurements by \textit{BeppoSAX} 
and \textit{RXTE} \cite{Guainazzi99,Lee99}. Thus the {\tt pexrav} cut-off energy was fixed at 150~keV, 
the reflection fraction was fixed at $\Omega/2\pi = 1$ (where $\Omega$
is the solid angle subtended by the reflecting material as seen by the primary source), and the iron
abundance was fixed at $1 \Zsun$. 

\begin{figure}
\psfig{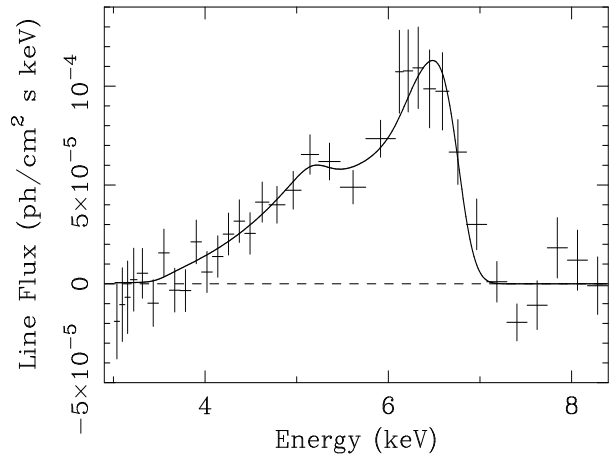}
\caption{The broad iron K$\alpha$ line from the 1999 long 
\textit{ASCA} observation of MCG--6-30-15.
The line has been constructed from a power-law 
plus reflection ({\tt pexrav}) fit to the underlying continuum. 
Plotted here is the total time-averaged SIS 
data, corrected for detector efficiency. Note the extremely broad, 
two-horned profile -- the FWHM of the blue core alone is $\sim$ 2--3 times
the detector resolution. The solid line shows 
the best-fit {\tt diskline} model for a Schwarzschild black hole, with 
$\chi^2 = 1885.8/1889$ \textit{dof} for the 
complete {\tt pexrav} plus {\tt diskline} model fit to SIS+GIS.
The best-fit {\tt diskline} model parameters 
are given in Table \ref{fkapar}.}
\label{fig_ironline}
\end{figure}

The model was fit simultaneously to the data from all four detectors between 3--10~keV. 
A good fit was obtained with $\chi^2 = 1885.8$ for 1889 degrees of 
freedom. The best-fit continuum power-law slope and normalization are $\Gamma = 2.082 \pm 0.009$ 
and $N_{3-10} = (1.90 \pm 0.03) \times 10^{-2}$ \phpspsqcm. The power-law slope 
is slightly higher than the slope $\Gamma_{97} = 1.94^{+0.06}_{-0.07}$ 
measured for the 1997 \cite{Iwasawa99} dataset. The difference in power-law slopes 
remains at the 2-$\sigma$ level even after systematic differences between 
the two analyses, such as the modification of {\tt pexrav} with {\tt rdblur} 
in this analysis and slightly different fitting procedures, are taken into 
account. 

\renewcommand{\arraystretch}{1.5}
\begin{table*}
\begin{tabular}{cccccccc}
\hline
Model & (1) & (2) & (3) & (4) & (5) & (6) & $\chi^2$/\textit{dof} \\
  & $\alpha$ & $\rin$ & $\rout$ & $i$ & $I$ & EW & \\
  &          & $\rg$  & $\rg$   & deg & \phpspsqcm & \eV & \\[5pt] 
\hline
\hline
Schwarzschild & $3.8^{+0.3}_{-0.3}$ & $6.00^{+0.04}_{-0.00}$ & $22^{+7}_{-3}$ & $34.1^{+0.4}_{-0.5}$ & $2.16^{+0.07}_{-0.08} \times 10^{-4}$ & $480^{+16}_{-18}$ & 1886/1889 \\
Extreme Kerr & $2.3^{+0.2}_{-0.1}$ & $1.8^{+0.1}_{-0.6}$ & $12.5^{+0.5}_{-0.8}$ & $37.7^{+0.8}_{-0.6}$ & $3.04^{+0.23}_{-0.12} \times 10^{-4}$ & $720^{+50}_{-30}$ & 1858/1889 \\
\hline
\end{tabular}
\caption{The best-fit parameters of the {\tt diskline} and {\tt laor} models for the iron-line profile
of the 1999 ASCA observation of MCG--6-30-15. (1) power-law index of the disk radial emissivity 
profile ($\epsilon \propto r^{-\alpha}$); (2) inner radius of the disk; (3) outer radius of the disk; (4) disk 
inclination angle; (5) iron-line intensity; (6) iron-line equivalent width. 
All quoted error bars are at the 
1-$\sigma$ level and include variations in the underlying continuum.}
\label{fkapar}
\end{table*}

The best-fit parameters for the {\tt diskline} model 
are shown in the first row of Table \ref{fkapar}, 
and the efficiency-corrected iron line 
profile is shown in Fig.\ \ref{fig_ironline}. 
Generally speaking, the {\tt diskline} 
parameters agree quite well with the parameters measured in 
the 1997 observation, confirming not only the presence of a broad iron K line 
in MCG--6-30-15, but also the properties of the accretion disk responsible for the line
emission. It is interesting to note, however, that while the 1997 data
preferred an inner radius of $6.7 \rg$, here $\rin$ pegs
at $6 \rg$, the radius of minimum stability for a non-rotating black hole and the lower bound
in the {\tt diskline} model. This, and the relatively high value of $\alpha$, suggest 
the presence of emission from within $6 \rg$, which is generally thought to be the signature of 
a rotating black hole.

To check for the significance of any emission from within $6 \rg$, 
a model for a maximally spinning ($a=0.998$) black hole was 
also fit to the data between 3--10~keV. 
The model is completely analogous to the one described above for 
a Schwarzschild black hole, with {\tt diskline} and {\tt rdblur} 
replaced with the {\sc xspec} model components {\tt laor} and {\tt kdblur}, which 
describe line emission and relativistic blurring from an accretion disk around an
extreme Kerr black hole \cite{Kojima91,Laor91}. Again, we obtained an 
excellent fit with $\Gamma = 2.096 \pm 0.009$ and 
$\chi^2 = 1858/1889$ \textit{dof}. The best-fit {\tt laor} parameters 
are shown in the second row of Table \ref{fkapar}. As indicated by the equally
acceptable values of $\chi^2$, the time-averaged line profile cannot constrain the
spin of MCG--6-30-15 given the current quality of the data.
It is worthwhile to note, however, that
in the best-fit Laor model, the entire line-emitting region of the accretion disk
shifts inward and stretches all the way down to $\sim 2$ gravitational radii, resulting
in a smaller emissivity index,
since it is no longer necessary to concentrate as much emission so 
close to the black hole. 

\section[Time-Resolved Spectral Features]
{Time-Resolved Spectral Features: Variability Studies}
\label{sec_tresspec}
\begin{figure*}
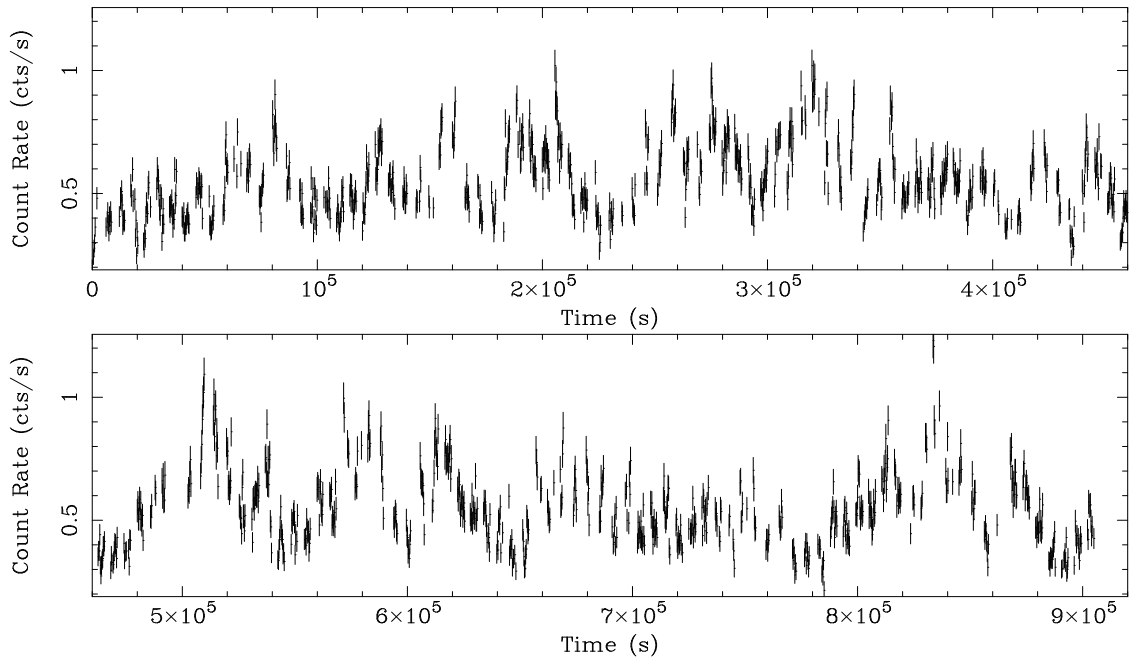

\centerline{\psfig{figure=sis.bgsub.3-10kev.256.lc.ps.t0-460000,angle=270,width=0.84\textwidth}}
\centerline{\psfig{figure=sis.bgsub.3-10kev.256.lc.ps.t460000-920000,angle=270,width=0.84\textwidth}}
\caption{The ASCA SIS light curve of MCG--6-30-15 in the 3--10~keV band. The 
data has been background-subtracted and is shown here in 256-s bins. The source 
was observed by ASCA from 19 July 1999 to 29 July 1999 for a total of 910 ks.}
\label{fig_lc_mcg63015}
\end{figure*}

The background-subtracted SIS 3--10~keV light curve for the 1999 observation 
is shown in Fig.\ \ref{fig_lc_mcg63015} with 256-s bins. As in 1997, the source 
appears to be highly variable, with the 3--10~keV count rate changing in the 
brightest flares by factors of 2--3 on timescales of less than a day. 

\subsection{Hardness Ratios}

\begin{figure}
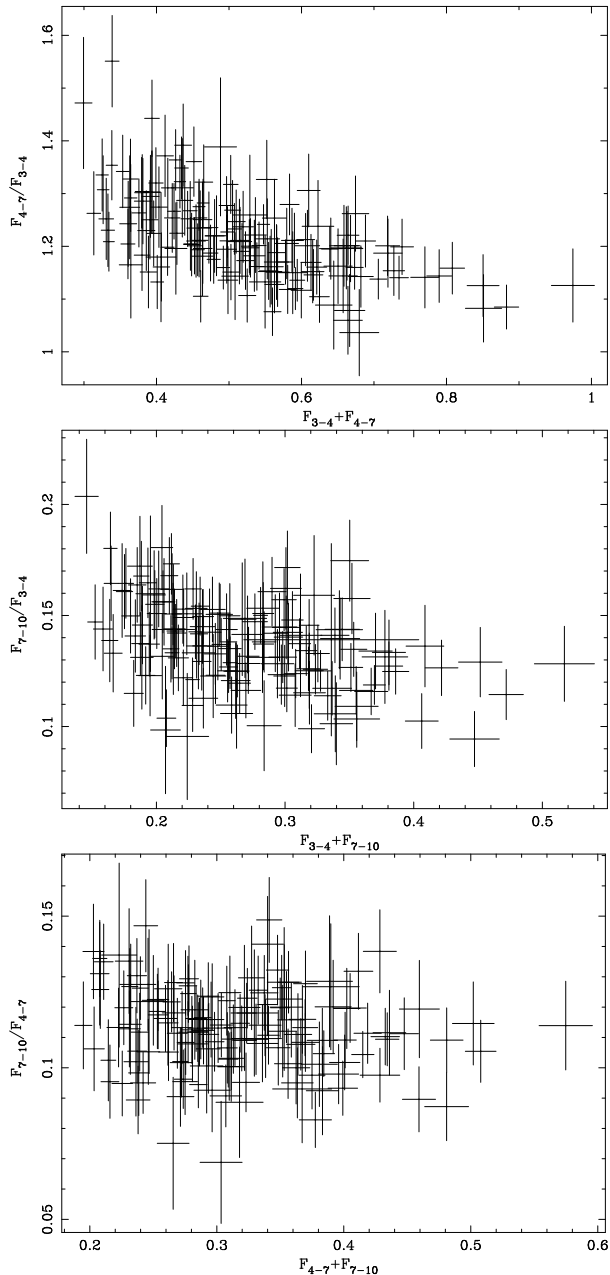

\psfig{figure=sis+gis.hrat.reb1.4-7_3-4.ps,width=0.45\textwidth,height=0.3164\textwidth,angle=270}
\psfig{figure=sis+gis.hrat.reb1.7-10_3-4.ps,width=0.45\textwidth,height=0.3164\textwidth,angle=270}
\psfig{figure=sis+gis.hrat.reb1.7-10_4-7.ps,width=0.45\textwidth,height=0.3164\textwidth,angle=270}
\caption{The hardness ratios $R_1$, $R_2$, and $R_3$ of MCG--6-30-15, plotted in 
separate panels from top to bottom against the total count rate in the relevant energy bands. 
The clear downwards trend seen in the first two hardness ratios indicate a definite spectral steepening as the source becomes brighter.}
\label{fig_4-7_3-4hrat}
\end{figure}

Hardness ratios provide a straightforward, ``zeroth-order'' means of investigating 
spectral variability. Being model-independent, they can provide direct insight 
into overall time-resolved spectral features. Also, because they do not require 
any spectral fitting, hardness ratios are free of any biases introduced by a 
specific fitting procedure. A total of three hardness ratios were examined: 
\begin{eqnarray*}
R_1 &=& \frac{F_{4-7}}{F_{3-4}}\\
R_2 &=& \frac{F_{7-10}}{F_{3-4}}\\
R_3 &=& \frac{F_{7-10}}{F_{4-7}}.
\end{eqnarray*}
The hardness ratios were calculated in each orbital period ($\sim 5700$ s) 
using the total SIS+GIS count rate in the relevant energy bands.
They are shown in Fig.\ \ref{fig_4-7_3-4hrat} plotted against count rate. 
$R_1$ and $R_2$ show a definite negative correlation with count rate 
($\chi^2 =$ 240.7 and 242.8, respectively, for 158 \textit{dof} against a 
constant hypothesis), indicating that the spectrum steepens as the source becomes 
brighter. On the other hand, $R_3$ does not appear to show much evidence for this 
correlation, although we note that any such correlation could be obscured by 
scatter due to the significantly poorer signal-to-noise of the $R_3$ band.

The hardness ratio trends shown here agree well with those found by Lee et al.\ (2000) in 
their recent analysis of the 1997 \textit{RXTE} observation, suggesting
that the observed correlation between spectral steepness and 
source brightness is an intrinsic and general feature of MCG--6-30-15. 
Since we expect the spectrum between 3--10~keV to be well-modeled by a simple 
power-law plus a relativistically broadened iron-line, the observed hardness 
ratio trends should correspond directly to a positive continuum index vs.\ 
count rate correlation. As we shall see in the next section, this is almost 
what is observed.

\subsection{The 12 Count-Rate Bins}
\begin{figure}
\centerline{\psfig{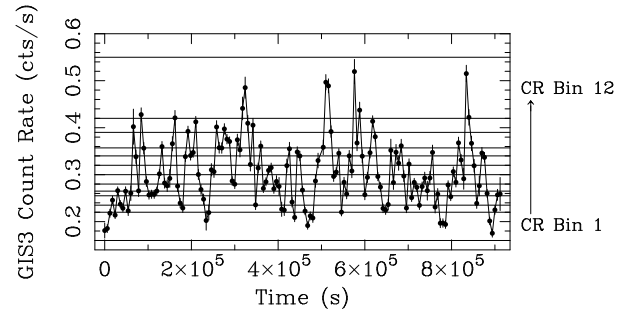}}
\caption{Essentially the same light curve as in Fig.\ \ref{fig_lc_mcg63015}, except derived
from the GIS3 detector and binned with 6000-s bins. Shown here are the 12 
horizontal slices of the GIS3 light curve that define the count-rate bins used for the
spectral variability analysis described in this paper.}
\label{fig_lc_crbin}
\end{figure}

The study of correlations between various spectral parameters and source flux was
carried out using twelve bins defined using \textit{horizontal} slices of 
the 3--10~keV light curve of the GIS3 detector. The procedure was as follows: 
the 3--10~keV GIS3 light curve was binned into
orbital periods ($6\times 10^3$~s). From this light curve, twelve
horizontal slices in count rate were chosen such that each slice contained approximately the 
same number of counts. These twelve horizontal slices 
are shown in Fig.\ \ref{fig_lc_crbin} along with 
the GIS3 light curve, binned into orbital periods.
The horizontal slices were then converted into
\textit{timing filters} (i.e. a union of disconnected vertical time slices, each with 
GIS3 count rate within the specified range). Finally, the timing filters were applied to the 
data from the other three detectors (SIS0, SIS1, GIS2) to produce the desired spectral files.

We decided to use horizontally sliced count-rate bins instead of the more
customary vertically sliced time bins for several reasons. 
First, binning in count rate allowed us 
to increase signal-to-noise while preserving the large amount of variability observed 
in the source. Time bins of comparable size would have smeared out most of the 
interesting flares and minima, thereby destroying the greater part of the correlations
to be described below. Secondly, using count-rate bins in some sense 
\textit{averages out} the dependence of $\Gamma$ and other spectral parameters 
on properties of the source other than count rate and focuses on the sole 
determination of these spectral parameters as a function of count rate.

\subsection{Photon Index vs.\ Count Rate Correlation}
\label{sec_indvsctrt}

The first correlation we investigated was the standard relation between
the continuum photon index ($\Gamma$) and count rate. A positive correlation 
between $\Gamma$ and source flux is believed to be a feature common to most Seyfert 1's 
(Perola et al.\ 1986, Nandra et al.\ 1991, Ptak et al.\ 1994; 
see also the review by Mushotzky et al.\ 1993).
Recent studies using high-quality data have firmly established the presence of 
a correlation in a number of AGN, including MCG--6-30-15 \cite{Vaughan2001}, 
NGC 5548 \cite{Chiang2000}, NGC 7496 \cite{Nandra2000},
and IC 4329A \cite{Done2000}. The 1999 \textit{ASCA} observation of MCG--6-30-15
is the longest and highest signal-to-noise observation of this source, and as such
it offers an unprecedented opportunity to determine the photon index as a function
of flux. However, the result of our analysis is surprising. Instead of finding
a simple positive correlation between $\Gamma$ and flux as expected from previous analyses, 
we report here a new trend: a flattening of $\Gamma$ at high count rate.

\begin{figure}
\psfig{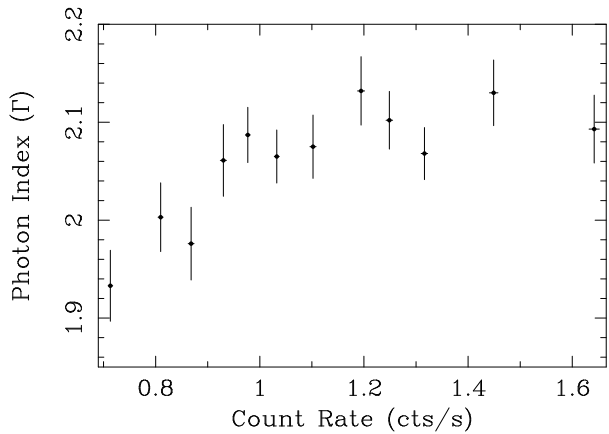}
\caption{The continuum photon index as a function of the 3--10~keV count rate. 
All errors shown are 1-$\sigma$. The photon index appears to saturate at high
flux to $\Gamma \sim 2.1$, a new feature of the $\Gamma$ vs. flux relation.}
\label{fig_indvsctrt}
\end{figure}

$\Gamma$ was determined using the blurred power-law plus 
reflection model described in Section \ref{sec_continuum_and_fka}. 
For simplicity, the iron line was not simultaneously determined in these fits, so as
to avoid any model degeneracies and biases 
(e.g.\ between the power-law slope and the iron-line flux) that might arise due to the reduced
signal-to-noise of the time-resolved data.
As in the time-averaged fit, the free parameters 
were $\Gamma$ and the continuum normalization. Fig.\ \ref{fig_indvsctrt} shows the 
best-fit photon indices plotted vs.\ the average 3--10~keV count rate of each flux bin. 
Absorption (warm or otherwise) is not expected to play a significant role above 3~keV in
this source (although we test this assumption below), 
and X-ray reprocessing features such as the iron line and the reflection
hump are estimated to contribute at most 15\% to the total 3--10~keV flux. Therefore
the 3--10~keV count rate should accurately reflect changes in the flux of the primary continuum. 
These changes are seen to result in a positive correlation of photon
index with flux when the flux is low, in qualitative agreement 
with previous analyses of MCG--6-30-15 and other sources. 
However at higher flux, the photon index gradually levels off to around 
$\Gamma \sim$ 2.1--2.2. A $\chi^2$-test against a 
linear correlation gives $\chi^2 = 15.3/10$ degrees of freedom, corresponding to a 
rejection of the linear correlation at 88\% confidence. 
The fit is dramatically improved with the power-law functional form 
$\Gamma = \Gamma_0-K F^{-\delta}$, which yields $\chi^2 = 6.65/9$ \textit{dof}
with a limiting photon index $\Gamma_0=2.12^{+0.12}_{-0.03}$ and a (negative) power-law index
of $\delta=3.9^{+2.0}_{-1.7}$. We believe this to be the first 
indication in an AGN of a saturation of the photon index at high flux. 
Should this result prove to be real, it could have serious implications for theoretical 
models of coronal activity and continuum production, as we shall discuss in 
Section \ref{sec_interpretation}.

We have tested the saturation of $\Gamma(F)$ at high flux and the result appears to be robust. 
Varying the fit prescription and frozen model 
parameters had little effect on the general trend. 
Similarly, it made little difference to the shape of the correlation whether relativistic 
blurring of the continuum ({\tt rdblur}) was included in the fit. 

Obvious from inspection of Fig.\ \ref{fig_indvsctrt} is the important role played by 
the highest flux bin ($F_{3-10} \approx 1.6$~cts/s) in determining the nature of the relation
between $\Gamma$ and flux. Excluding the highest flux bin does improve the
linear fit to a marginally acceptable $\chi^2 = 11.5/9$ \textit{dof}. However, the residuals exhibit
additional structure that still suggest a saturating $\Gamma$. This is confirmed by the
fact that fitting with the power-law functional form results in a best-fit nearly identical to
the full 12-bin fit. The stability of the power-law fit, along with the fact
we can think of no good physical reason why the highest flux bin should be excluded from the relation,
give us confidence that the apparent saturation of $\Gamma$ is not being caused by a single bin.

Other effects, such as correlated errors and model-specific biases, could be more subtle. 
MCG--6-30-15 is known to contain a highly variable warm absorber, with the absorption dominated
by O~{\sc vii} and O~{\sc viii} edges at 0.74 and 0.87~keV \cite{Fabian94,Lee2001}. While the warm absorber is
not expected to play a significant role above 3~keV, we have nonetheless tested for the possibility
that variable warm absorption could have an effect on the $\Gamma$ vs. flux relation.
Severe detector degradation below 2~keV due to radiation damage prevented detailed and realistic 
warm absorber fits from being done in the 1999 observation. Instead, we attempted to model the effect
of warm absorption above 3~keV using an additional cold absorption model component ({\tt wabs}, Morrison \& McCammon (1983)). The fits were redone with the column density 
of this extra {\tt wabs} left as a free parameter. The fits actually preferred a rather high column density
of $\sim 10^{22}$ cm$^{-2}$, which resulted in slightly softer photon indices. However, since the
amount of absorption was not correlated with flux, the resulting $\Gamma$ vs.\ flux relation 
was largely unchanged and still showed the saturation of $\Gamma$ at high flux. 

A systematic study was conducted to ascertain whether there could be any 
count-rate dependent bias in the determination of $\Gamma$.\footnote{
Strictly speaking, it is not the count rate, but rather the total
number of counts (i.e.\ the signal-to-noise level) in each bin that
determines the level of bias in the fit.
Since the count-rate bins were defined so as to contain approximately the same number of counts,
such a bias should in principle have a minimal effect on the shape of the trend.}
10 simulated spectra were produced in each count-rate bin using the best-fit photon index and 
continuum normalization for that bin. Exposure times, response curves, and 
background spectra were derived from real data so as to mimic as closely 
as possible the actual fitting process. $\Gamma$ was then measured 
for the 10 simulated spectra in each count-rate bin, and the average $\Gamma$
was found to differ from the input photon index by 
less than 1\% in every count-rate 
bin (well within the 1-$\sigma$ errors on the actual $\Gamma$), with no
systematic trend with respect to count rate. Thus the saturation of $\Gamma$ at high flux
cannot be accounted for by a flux-dependent bias in the fitting process.

Finally, a similar analysis was carried out to measure the continuum photon index as a function of 
flux during the 1997 \textit{ASCA} observation of MCG--6-30-15. The result agrees
remarkably well with that of the present observation, providing further evidence (along with
the hardness ratios) that the $\Gamma$ vs.\ flux correlation is a general feature of 
MCG--6-30-15. However, because the source was approximately 20\% fainter in 1997, 
the bulk of $\Gamma(F)$ for that observation falls below the saturation regime of
$\approxgt$ 1 ct/s. Indeed, of the three long \textit{ASCA} observations of MCG--6-30-15
in 1994, 1997, and 1999, only the latter provides the exposure time and average source
luminosity necessary to adequately probe the saturation regime of the $\Gamma$ vs. flux relation. 

\subsection{Iron-line Variability}
\label{sec_fkavar}

Temporal variations of the iron-line profile of MCG--6-30-15 in the three long 
\textit{ASCA} observations (1994, 1997, and 1999) were previously studied by Matsumoto et al.\ (2001). 
They showed using the normalized RMS variability spectrum of 
the 1999 data that the variability of the line band (5--6.6~keV) decreased faster than that of the continuum
band as one moved to longer timescales. This suggests that the line varies on a shorter characterstic timescale
than the continuum.
In addition, using double gaussian fits to the red and blue wings of the broad line, they found the line flux to be
significantly variable during the 1994 and 1997 observations, but not during the 1999 observation. The
line equivalent width, however, was variable in all three. The equivalent widths of the red and blue wings
showed a very weak positive correlation, with an unacceptably large scatter around a linear best-fit. The
total equivalent width tended to be anti-correlated with continuum flux, contrary to the constant behavior 
that would be expected from the simplest coronal disk-line models. 

In this analysis, we have studied in greater detail the relationship, if any, between properties of the iron line
and the continuum flux. Detailed fits in the 3--10~keV band using the Laor disk-line plus continuum 
model described in Section \ref{sec_continuum_and_fka} were performed for each of the twelve 
count-rate bins. The free parameters were the 
photon index, continuum normalization, inner and outer disk radius, and total line flux. The inclination angle
of the disk was frozen at its time-averaged best-fit value (see Table \ref{fkapar}), and the rest-energy
of the line was fixed at 6.4~keV. The disk emissivity index was also frozen at its time-averaged value, since
simultaneous determination of the emissivity index and the disk radii was subject to large degeneracies
due to the reduced signal-to-noise of the time-resolved spectra.

\begin{figure*}
\centerline{\psfig{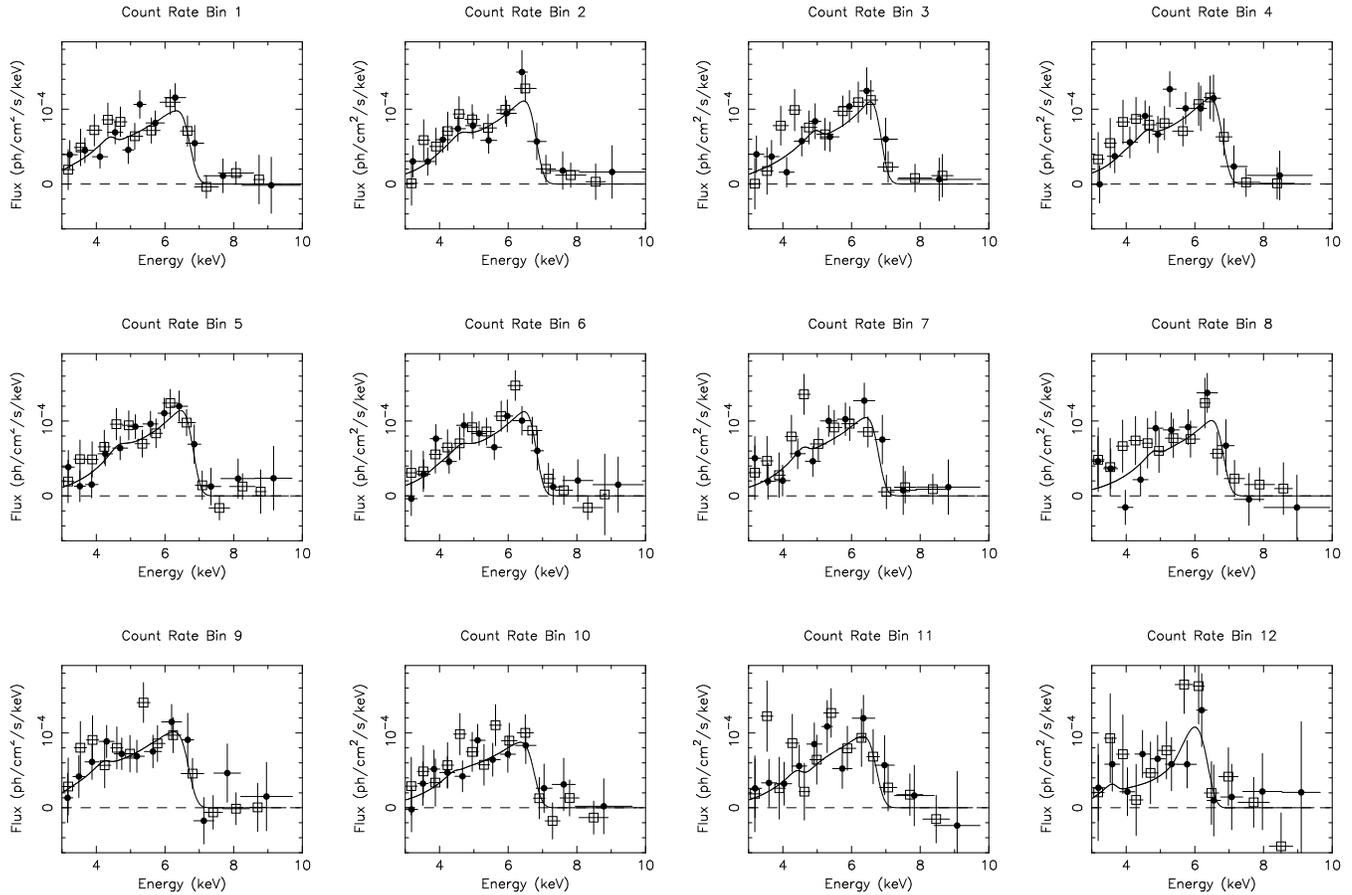}}
\caption{The broad iron line in twelve count rate bins, along with the best-fitting Laor model. Filled circles 
(open squares) correspond to data from the SIS (GIS). The bins are numbered in order of increasing flux.}
\label{fig_fkavar_spec}
\end{figure*}

Fig.\ \ref{fig_fkavar_spec} shows the efficiency-corrected iron line profiles obtained in each of the 
12 count-rate bins, along with the best-fitting model in each case.
The energy and flux scales are the same in each panel. Even with the reduced signal-to-noise of the 
time-resolved spectra, a very broad iron line can still be clearly resolved in nearly every bin. In most
cases, an extended red tail is seen stretching all the way down to (or even below) 3~keV, suggesting the
presence of line-emitting matter extremely close ($r \sim$ a few $\rg$) to the central black hole at almost
all levels of source brightness. Overall, the line appears to be remarkably
stable across the flux bins, with the exception of perhaps the highest-flux bin, where the red wing seems
to weaken considerably while the narrower blue wing remains at $\sim$ 6~keV.

\begin{figure*}
\centerline{\psfig{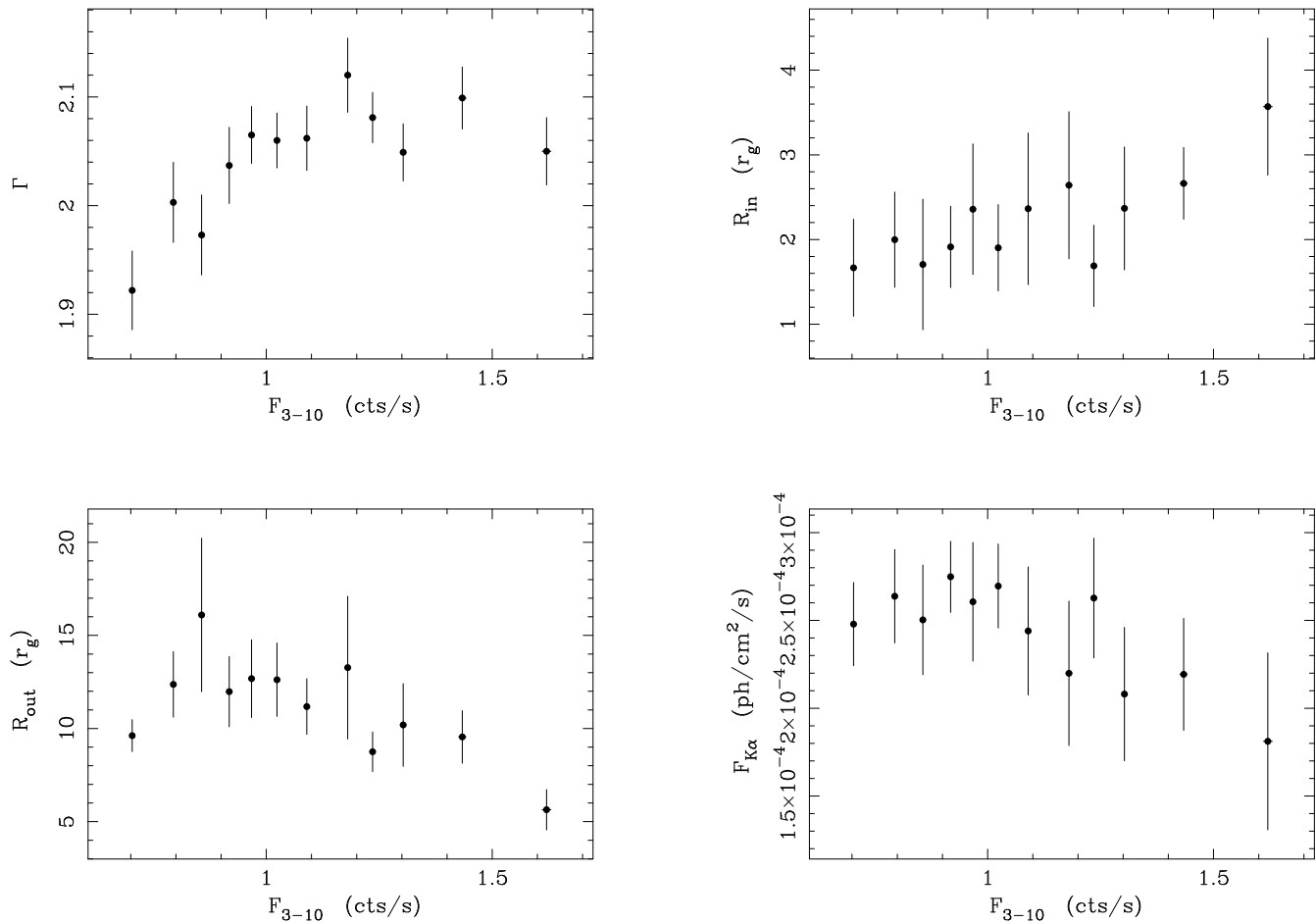}}
\caption{The best-fit parameters (continuum photon index, inner/outer disk radius, iron-line flux) of
the iron-line variability fits, plotted vs.\ 3--10~keV count rate.}
\label{fig_fkavar_par}
\end{figure*}

The spectral parameters shown in Fig.\ \ref{fig_fkavar_par} offer a more precise picture of the 
variability of the line with flux. The continuum photon index $\Gamma$ has been plotted in addition to the 
inner/outer radii and the line flux in order to demonstrate that the $\Gamma$ vs.\ flux relation 
found in Section \ref{sec_indvsctrt} holds even when the iron-line and continuum are simultaneously fit. 
Regarding the line parameters, we see that they accurately reflect the apparent stability 
that was seen in the line profiles of Fig.\ \ref{fig_fkavar_spec} (excepting the highest-flux bin). 
At every flux level, $\rin$ is within $6 \rg$, the radius of marginal stability for a 
Schwarzschild black hole. 
More importantly in the context of coronal models, the fits show that the iron-line flux 
clearly does not increase with the continuum flux. 
The iron-line flux is consistent with a constant level 
($\chi^2 = 7.2$ for 11 d.o.f.) over nearly a factor of three change in source count rate. 
The lack of any positive correlation between line and continuum flux
appears to be robust, confirming the previous findings of Lee et al.\ (2000) and Vaughan \& Edelson (2001) for MCG--6-30-15.

The uncorrelated line and continuum fluxes present a serious problem for the simplest coronal models
in which the iron line is produced by reflection of the primary continuum off a cold accretion disk, 
with the continuum generated in a corona directly above the disk. 
In such a scenario the iron line flux would be expected to respond proportionately to 
variations in the primary continuum, with the equivalent width remaining roughly constant with respect
to the continuum flux. Our findings clearly
indicate that the equivalent width of the iron line actually \textit{decreases} with flux in MCG--6-30-15.
Although a positive correlation between line flux and continuum flux has been observed in some AGN, 
such as NGC 7314 \cite{Yaqoob96}, it appears to be completely lacking in other AGN besides MCG--6-30-15,
most notably in NGC 3516 \cite{Nandra99} and
NGC 5548 \cite{Chiang2000}. The fact that the line and continuum flux are not 
correlated in some AGN indicates that the reflection process, at least in these AGN, is more complex 
than that described by a simple coronal model. 

Stronger evidence for variation with flux is found for the outer radius $\rout$. Since the
disk inclination angle has been fixed to its best-fit time-averaged value, $\rout$ is relatively
well-constrained by the well-defined peak and maximum energy of the blue wing. A constant level
is a poor description of $\rout$ vs.\ count rate, with $\chi^2 = 27.3/11$ \textit{dof}. The fit improves
significantly when one fits with a linear relation ($\chi^2 = 16.4/10$ \textit{dof}, best-fit slope $-4.4 \pm 1.3$).
Excluding the lowest-flux bin leaves the quality of the constant fit virtually unchanged, while the linear
fit improves further to $\chi^2 = 4.8/9$ \textit{dof} with slope $-8.9 \pm 1.9$. Thus, overall, the outer radius
appears to exhibit a significant anticorrelation with source flux, suggesting that the illumination 
pattern on the disk may shift closer to the black hole when the source brightens. 
It is intriguing that $\rin$ also appears to exhibit a slight positive
correlation with flux, although it is not statistically significant. The trend, if present, 
along with the anticorrelation of $\rout$ would imply that the line emission
originates in a more localized part of the disk when the source is bright. One might 
expect this sort of behavior if variations in source flux were due to flaring active regions on the disk. 
The overall trends in $\rin$ and $\rout$ could then be understood
by assuming that the dominant active region was relatively stable throughout the 1999 observation 
and was located around 4--5 $\rg$.

A detailed, exhaustive analysis of the many possible degeneracies between various
fit parameters is beyond the scope of this paper. Thus we cannot rule out the possibility 
that the trends shown here are simply a result of correlated errors. Nevertheless, we were
able to rule out one of the more obvious possible correlated errors with the following simple check.
The plot of line flux vs.\ source flux in Fig.\ \ref{fig_fkavar_par} strongly suggests an anti-correlation
between line and source flux. Suppose however that the line flux was actually \textit{constant}. Then
since the count-rate bins contain approximately the same number of continuum counts, 
the higher count-rate bins, being shorter in duration, would contain less counts from the iron line.
Accurate determination of the line parameters would be more difficult, and in particular, the wings
of the broadened line would be harder to detect. This would bias the line flux downwards and in addition
spuriously narrow the line emitting region. To check that this systematic 
error due to decreased line counts is not the source of the behavior shown in Fig.\ \ref{fig_fkavar_par}, 
we have binned together the last two count-rate bins, and the spectral fits have been redone 
for this summed spectrum. The spectral parameters obtained for the 
summed spectrum follow the general trend, and they are found to lie between those of the 
two last count-rate bins, indicating that at least part of the variation in spectral parameters 
cannot be explained by a simple decrease in line counts, with the line flux constant.

\section[Interpretation]
{Interpreting the $\Gamma$ vs.\ Flux Correlation}
\label{sec_interpretation}

\subsection{The Two Power-Law Model}
\label{sec_mo2po}

\begin{figure*}
\psfig{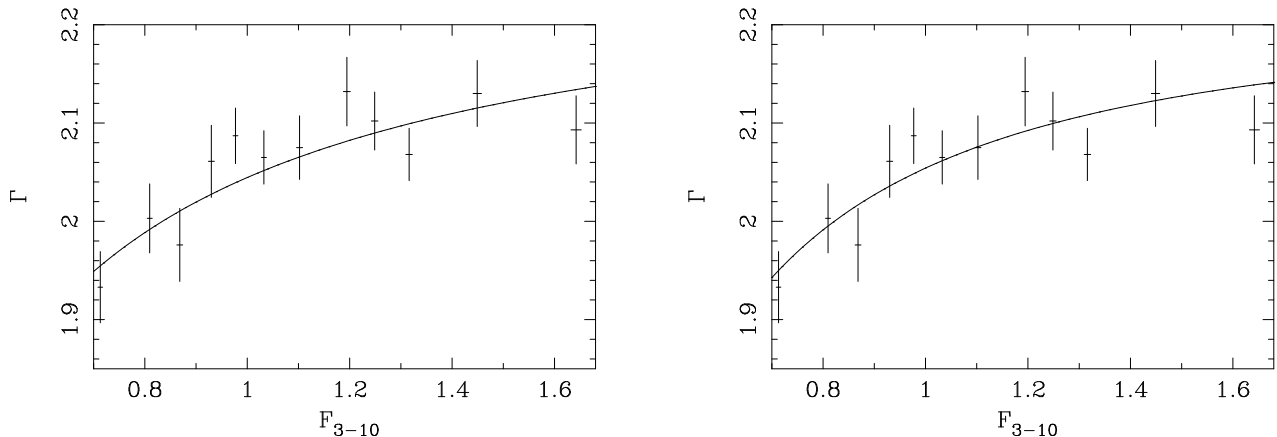}
\caption{Good fits to the data for the two power-law model (left) with $\Gamma_1 = 1.85$, $\Gamma_2 = 2.25$,
$F_1$ = 0.54; and the thundercloud model (right) with $\tau = 1.5$, $C = 0.001$, and $D=1.4$. For both models
the predicted $\Gamma$ vs.\ flux relations were interpolated using the function $\Gamma(F) = \Gamma_0 - K F^{-\delta}$.
As one can see from the plots here, it is impossible to rule out either model from the $\Gamma$ vs.\ flux relation
alone, although the thundercloud model certainly has the advantage of being more theoretically motivated.}
\label{fig_mo_2po_thund}
\end{figure*}

We propose two different models that can simultaneously account for the observed 
saturation of the photon index at high flux and the lack of correlation between the iron-line and continuum
fluxes. The first is an ad hoc 
phenomenological model consisting of two power-law continua, 
one constant with flux and photon index $(F_1,\Gamma_1)$, and the other variable
with $(F_2,\Gamma_2)$. To test this model,
a coarse grid in $\Gamma_1$ and $\Gamma_2$ was produced with $\Gamma_1$ ranging from 1.75 to 2.0 and
$\Gamma_2$ from 2.2 to 2.3. $F_1$ was fixed at 0.54 cts/s (chosen to
correspond approximately to the minimum of the 3--10~keV light curve), 
because $\Gamma_1$ and $F_1$ proved to be highly degenerate parameters.
Simulated spectra for each value of $\Gamma_1$ and $\Gamma_2$ were produced in every count-rate bin and
were folded through the exposure times, response matrices, and effective areas of the actual detectors. 
Normalizations were determined in each bin by requiring the overall 3--10~keV count-rates of the two
power laws to equal $F_1$ and $F_2 = F-F_1$ respectively, where $F$ was the 
average count rate of the actual data in that bin.
The simulated spectra were fit with a single power-law model between 3--10~keV to produce $\Gamma$ vs.
flux relations for each value of $\Gamma_1$ and $\Gamma_2$. Shown in the left panel of Fig.\ \ref{fig_mo_2po_thund}
along with the data is the ``best-fit'' for our coarse grid 
($\Gamma_1 = 1.85$, $\Gamma_2 = 2.25$ with $F_1=0.54$; $\chi^2 = 10.5/10$ \textit{dof}),
interpolated using the functional form $\Gamma = \Gamma_0-K F^{-\delta}$.
The term ``best-fit'' is slightly misleading, however: for every value of $\Gamma_2$, there 
was a value of $\Gamma_1$ that yielded an acceptable fit. 
In any event, the two power-law model, although poorly constrained by the data, 
is clearly capable of reproducing well the observed form of $\Gamma(F)$.

Aside from the fact that it models well the measured $\Gamma$ vs.\ flux relation, the two power-law model
is attractive because it offers a chance of decoupling the iron-line and continuum variability.
Previous analyses, and the detailed results of Section \ref{sec_fkavar}, indicate strongly
that the iron line flux is uncorrelated with the continuum flux in MCG--6-30-15. As we discussed
above, this is very puzzling in the context of simple coronal models in which the line is expected
to respond directly to changes in the primary continuum. The two 
power-law model offers a potential solution to this problem: by associating
the iron line with reprocessed radiation from the constant component alone, the continuum
is free to vary in response to the variable component without having any effect on the iron line.
Although the iron line is known to exhibit significant temporal variability \cite{Iwasawa99,Matsumoto2001}, we see from Section \ref{sec_fkavar} that it is essentially stable 
across flux bins. Therefore we must consider the possibility that the ``constant'' 
$F_1$ component varies to some extent as well, but that its variations are smeared out 
in the $\sim 80$ ks count-rate bins. This would not change the $\Gamma$ vs. flux relation predicted by the model.

At present the two power-law model is basically a phenomenological model, with very little theoretical motivation.
We suggest a few possible explanations for the origin of the two power-law components. 
One possibility is that the ``variable'' $F_2$ power-law component originates in the inner edge of the accretion 
disk, where the solid angle subtended by the reflecting material is the least; 
while the ``constant'' $F_1$ power-law is produced farther out (but still in the inner part of the disk).  
$F_2$ could here be associated with magnetized accretion just inside the marginally stable
orbit \cite{Krolik99}, as simulations \cite{Hawley2001} show that this process can be very noisy.
Another possible scenario is that flares occurring at different heights above the corona are responsible for the
two power laws. DiMatteo et al.\ (1999) showed that the spectral properties of the 
soft-state and hard-state of the GBHC GX 339-4 could be qualitatively explained by flares occuring close
to and far above the accretion disk, respectively. Moreover, they found that the ionization state of the disk
was quite different depending on the height of the flares, with flares closer to the disk naturally ionizing
it more. Therefore associating the $F_1$ and $F_2$ components with flares high above and close to the disk,
respectively, would not only account for the difference in continuum shape, but also, if the flares close
to the disk completely ionized away the iron-line, would explain the lack of an iron line accompanying $F_2$.

Finally, there is the possibility that a small but non-negligible non-thermal component is present in the corona.
A hybrid thermal/non-thermal corona has been explored as a model for the soft-state Cygnus X-1 X-ray spectrum
and was found to explain well its overall shape \cite{Gierlinski99}. Single scattering of 
disk seed photons off the non-thermal component gives rise to a hard, high-energy power law, while
multiple scattering off the thermal part produces a softer power-law tail above the thermal disk 
emission at lower energies.
A similar situation in MCG--6-30-15 would lead to the harder $F_1$ (softer $F_2$)
component of our two power-law model being associated with the non-thermal (thermal) part 
of the corona, and with the two states varying independently of one another. An obvious problem with
this explanation is that strong evidence for a thermal corona in MCG--6-30-15 
is provided by observations of a cutoff in the high energy X-ray spectrum \cite{Guainazzi99,Lee99}.
We note, however, that even small non-thermal tails in the coronal 
electron distribution can apparently have sizable effects on the shape and flux of the Comptonized spectrum 
\cite{Wardzinski2001}. How well the observations limit the size of any non-thermal tails in the corona
of MCG--6-30-15, as well as what affect these might have on the 3--10~keV X-ray spectrum, remains to be seen.

\subsection{The Thundercloud Model}

A more theoretically motivated model of coronal variability that predicts quite generally 
the saturation of the continuum photon index at high flux was recently proposed by Merloni \& Fabian (2001). Their
so-called ``thundercloud model'' was based on the idea that active regions in the corona could, under the right
conditions, trigger avalanches of neighboring flares, giving rise to progressively larger
``magnetic thunderclouds'' that are responsible for inverse Compton scattering soft disk photons into the observed X-ray continuum. Larger active regions tend
to be more luminous (since they contain more magnetic reconnection sites) and produce
softer spectra. Therefore when the source is bright, the continuum is dominated by soft emission from large
active regions. During the largest flares, the active regions approach the limiting case of a slab-like 
geometry covering the entire disk, resulting in a saturation of $\Gamma$ at high flux.

In the thundercloud model, the luminosity of an active region is assumed to scale with its size 
via the relation $L(r) \propto r^D$, where $D$ is a free parameter 
of the model which describes both the internal structure of the region 
(a kind of fractal dimension) as well as any radial dependence there might be in the input radiation from the 
accretion disk. The number of active regions of size $r$ generated at time $t$ is chosen from a Poissonian
distribution with mean $n(r,t)$, which in turn is determined by requiring that the corona be in a stationary state. 
This gives the relation $n(r) = n(r,t)t_0(r)$, where 
$n(r) \propto r^{-p}$ is total number of active regions of size $r$
and $t_0(r)$ is the typical lifetime of these regions. Here $p$ is given by the relation $p = 2D+3-\gamma$, where
$\gamma$ is the slope of the red noise part of the Power Density Spectrum (PDS). For comparison with MCG--6-30-15, 
the slope of the PDS was fixed at $\gamma = 1.5$, in agreement with the literature \cite{Nowak2000,Matsumoto2001}.
In addition to $D$, the other two free parameters of the model were the covering fraction $C$, 
which determines the total number of active regions; and the optical depth $\tau$ of the corona. The mass of
the black hole, accretion rate, and maximum/minimum flare size were fixed at $M_{\rm BH} = 10^7 \Msun$, $\dot{m} = 0.1$,
$r_{\rm min} = 0.02$, and $r_{\rm max} = 4$, respectively. These serve mainly to set the overall luminosity and time
scales.

The $\Gamma$ vs.\ flux relation predicted by the thundercloud model was compared to the data using
a grid with optical depth $0.5 \le \tau \le 2$, covering fraction 
$1\times 10^{-4} \le C \le 2\times 10^{-3}$, and luminosity scaling index $0.6 \le D \le 1.6$. In addition
to the constraint given by the $\Gamma$ vs.\ flux relation, the overall normalized RMS variance of the model
light curve was required to lie within the range 20--30\% to match the that of MCG--6-30-15. Shown in the
right panel of Fig.\ \ref{fig_mo_2po_thund} is a 
particularly good description of the data with $\tau = 1.5$, $C = 0.001$,
and $D = 1.4$ (the RMS variance was 21\%). Overall, the coronal optical depth
was the most well-constrained parameter, being required to lie in the range $1 \le \tau \le 2$. The constraint
on $\tau$ is a consequence of the fact that it determines the limiting value of the photon index:
it is essentially the only remaining free parameter at the highest flux levels when the active regions 
limit to a disk-covering, slab-like geometry. 

It is worth considering briefly the physical implications of $\tau > 1$ for iron-line variability. 
Since a $\tau > 1$ corona tends to smother any reflected radiation from the disk,
the line should essentially vanish during higher flux states, 
when the thunderclouds cover most of the accretion disk. Meanwhile, in
lower flux states, the more or less continuous series of small, localized flares should lead to a roughly constant
line. Referring back to the time-resolved line profiles of Fig.\ \ref{fig_fkavar_spec} 
and the iron-line vs.\ continuum flux plot of Fig.\ \ref{fig_fkavar_par}, it is intriguing to
see that, within the limitations of the poor signal-to-noise, the behavior of the iron-line profile
as a function of flux does indeed appear to support this prediction of the thundercloud model.

The covering fraction and scaling index were much more difficult parameters
to constrain, as they contribute principally to determining the overall curvature of $\Gamma(F)$, 
which is poorly measured by the present dataset. We can, however, limit these parameters to
$C > 0.01$\% and $D < 2$. These bounds are related, since values of $C$ below $0.01$\% require 
$D > 2$ (and vice versa) in order to achieve realistic values of the RMS variance. This can be understood by noting
that when the total number of active regions is small, their luminosity must grow quickly with their size
in order to produce significant variations in overall luminosity over time. Such large values of $D$ in turn require $p$
to be large, which gives a corona dominated by very small active regions. This results in a
rather unrealistic light curve consisting of long periods of stability due to small active regions, punctuated
with the occasional huge flare from a large active region.

\section[Summary]
{Summary}
\label{sec_summary}

An analysis of the spectral variability of MCG--6-30-15 during the 1999 
long \textit{ASCA} observation has been presented. The principal findings of the analysis
will now be summarized. A model consisting of a power-law continuum, plus reflection
and a relativistically broadened iron-line from the accretion disk, gave a good fit to the 
time-averaged 3--10~keV spectrum. The spectral parameters of 
the broad iron line were generally in good agreement with the results of previous analyses of 
MCG--6-30-15. While not statistically significant, the data seemed to prefer an 
inner radius of the disk within the radius of marginal stability $\rms$ for 
a Schwarzschild black hole, in contrast to earlier observations which found a best fit 
inner radius just outside $\rms$.

A study of the time-resolved iron-line profile revealed 
a complete lack of any positive correlation between iron-line flux and 
continuum flux, confirming the results of
previous analyses. In addition, marginal evidence was found for a narrowing with flux 
of the line-emitting region on the accretion disk, suggesting that iron-line variability may be due in part to 
changes in the illumination pattern of the disk.

The power-law continuum was found to exhibit a positive correlation between photon
index and source flux that \textit{levelled off} at high flux, with $\Gamma$ limiting around 2.1--2.2.
Two different models were proposed to explain the observed 
saturation in $\Gamma$, and both were capable of describing
the trend well. The first model proposed was a simple phenomenological 
model consisting of two power laws, one constant with $\Gamma \sim 1.8$, and the variable with $\Gamma \sim 2.2$.
It was also suggested that the lack of correlation between iron-line and continuum flux could be explained
with this model by associating the line production with the harder, constant component.
Possible origins for the two distinct power laws were considered, and coronal 
flares of varying heights above the disk and a small non-thermal component in the corona were put forth
as viable hypotheses.
The second model advanced to explain the $\Gamma$ vs. flux relation was the so-called ``thundercloud model'' 
of coronal variability recently proposed by Merloni \& Fabian (2001). The thundercloud model also 
generally predicts a saturation in $\Gamma$ at high flux as the growing thunderclouds 
approach a slab-like geometry above the accretion disk. A detailed comparison with the 
thundercloud model constrained the optical depth of the corona to lie in the range $\tau \sim $1--2,
and a lower limit on the coronal covering fraction of $C \ge 0.01$\% was obtained. In addition,
the thundercloud model with $\tau \sim $1--2 predicts a pattern of iron-line variability (roughly constant
line profile when the continuum flux is low; decreasing line flux when the continuum flux is high) 
which accords well with the observed behavior of the line.

\section*{Acknowledgements}
We thank Andrea Merloni for his assistance with the thundercloud model and Simon Vaughan for 
many fruitful discussions. DCS thanks the Herchel Smith Fellowship for support. KI 
thanks PPARC for support. ACF thanks the Royal Society for support.

\end{document}